\def\year{2020}\relax
\documentclass[letterpaper]{article} 
\usepackage{aaai20}  
\usepackage{times}  
\usepackage{helvet} 
\usepackage{courier}  
\usepackage[hyphens]{url}  
\usepackage{graphicx} 
\def\UrlFont{\rm}  
\usepackage{graphicx}  
\title{Towards Automated Sexual Violence Report Tracking}
\author{
Naeemul Hassan\textsuperscript{\rm 1}, Amrit Poudel\textsuperscript{\rm 2}, Jason Hale\textsuperscript{\rm 2}, Claire Hubacek\textsuperscript{\rm 2}, Khandakar Tasnim Huq\textsuperscript{\rm 3} \\
\Large{\textbf{Shubhra Kanti Karmaker Santu\textsuperscript{\rm 4}, Syed Ishtiaque Ahmed\textsuperscript{\rm 5}}} \\
\textsuperscript{\rm 1}University of Maryland, College Park, \textsuperscript{\rm 2}University of Mississippi, \textsuperscript{\rm 3}Khulna University of Engineering and Technology \\
\textsuperscript{\rm 4}Massachusetts Institute of Technology, \textsuperscript{\rm 5}University of Toronto
}

\usepackage{booktabs}
\usepackage{caption}
\usepackage{subcaption}
\usepackage{enumitem}
\usepackage{multirow}

\usepackage{color}

\newcommand\tab[1][0.75cm]{\hspace*{#1}}

\begin{document}

\maketitle



\begin{abstract}
Tracking sexual violence is a challenging task. In this paper, we present a supervised learning-based automated sexual violence report tracking model that is more scalable, and reliable than its crowdsource based counterparts. We define the sexual violence report tracking problem by considering victim, perpetrator contexts and the nature of the violence. We find that our model could identify sexual violence reports with a precision and recall of $80.4\%$ and $83.4\%$, respectively. Moreover, we also applied the model during and after the \#MeToo movement.
Several interesting findings are discovered which are not easily identifiable from a shallow analysis.
\end{abstract}

\section{Introduction}
According to World Health Organization (WHO), sexual violence is defined as any sexual act, attempt to obtain a sexual act, unwanted sexual comments or advances, or acts to traffic, or otherwise directed, against a person's sexuality using coercion, by any person regardless of their relationship to the victim, in any setting, including but not limited to home and work~\cite{krug2002world}. This is a serious offence that causes several negative impacts on the victims' physical, psychological, and social health~\cite{fitzgerald1997antecedents,loy1984extent}. Unfortunately, the prevalence of sexual violence is very high across the globe~\cite{gelfand1995structure}, and women are the principle victims of sexual violence. According to UN Women and World health Organization (WHO), $35\%$ (more than one in every three) of women worldwide have experienced either physical and/or sexual intimate partner violence or sexual violence by a non-partner at some point in their lives~\cite{unwomen}.

Despite its disturbing prevalence and severity, it is often difficult to detect sexual violence and reach out to a victim in a timely manner~\cite{aroussi2011women}. This challenge mostly comes from the fact that sexual violence is often subjective and context-dependent~\cite{clay2003context}. Without a report by the victim, many sexual violence acts are hard to recognize. However, self-reporting is not always easy for the victims because of various social and cultural constraints including taboos around discussing sex, stigma around the victims of sexual violence, and the possibility of victim blaming in many societies around the world~\cite{fisher2003reporting,cdc,spencer2017sexual}. These challenges constitute a 'silence' that surrounds sexual violence and that has long been impeding the process of reducing the occurrence and impact of this problem.

While the victims tend not to self-report through traditional mechanisms such as filing a police report~\cite{felson2005reporting} or going to corresponding workplace/educational system (e.g., Title IX)~\cite{cantor2015report} due to uncertainty and complexity of the process among other reasons~\cite{krebs2007campus}, they often take their frustration into social media and disclose their violent experience to friends, peers, or followers~\cite{andalibi2016understanding}. The objective of this paper is to identify the sexual violence victims by mining self-reports from social media.

In October, 2017, an online movement began over social media where women from all over the world started sharing their stories of experiencing sexual violence using the hashtag \#MeToo. This online movement was fueled up by actress Alyssa Milano's invitation to all women to reveal the name of the perpetrators of sexual violence. She said, ``If all the women who have been sexually harassed or assaulted wrote `Me too' as a status, we might give people a sense of the magnitude of the problem'', on October 15, 2017.
After Milano's tweet, within 24 hours, there were more than 4.5 million posts on different social network platforms with hashtag \#MeToo (or some variants of it).


Thus, the \#MeToo movement created an unprecedented opportunity for women to express their untold feelings to the world, and for researchers and practitioners to understand various patterns of violence in various contexts. Millions of \#MeToo posts allow us learning the kind of violence, victim's relationship with the perpetrator, location and time of the incident, and victim's emotional response among others. This, in turn, also creates the opportunity to identify if a written statement expresses a report of sexual violence, even if it's not explicitly tagged with \#MeToo. This capability is particularly important in order to automatically detect violence from a massive stream of text data over social media. An automated system for identifying posts that report sexual violence will eliminate the need for a huge amount of human power to observe and evaluate each post separately. Furthermore, attention to posts on social media often depends on the social capital of the poster, which shrinks the way for victims with small capital to get proper attention. An automated system can eliminate such limitations and provide a scalable platform to instantly detect a post with a report of violence and inform concerned parties for reacting. Such automated tools can also be used for understanding the prevalence, severity, and nature of violence in different geographic locations in different times of the year.  

To this end, we have developed a data-driven, AI based automated sexual violence report tracking model that can \textbf{i)} detect a sexual violence report, \textbf{ii)} characterize the report with respect to several contexts-- \emph{gender of the victim}, \emph{nature of the violence in terms of severity}, \emph{category of the perpetrator}, and \textbf{iii)} automatically tag the perpetrator describing portion from the report. Our model can detect the sexual violence reports with a precision and recall of $80.4\%$ and $83.4\%$, respectively. It can also annotate the perpetrator identity from the report with an average F1-measure of $76\%$. Overall, we make the following contributions in this paper-

\begin{enumerate}
    \item We gather about a million \#MeToo hashtagged tweets, systematically process the corpus to remove noise, and identify potential sexual violence reports using Semantic Role Labeling (Section~\ref{sec:data-process}). A subset of the corpus is made public for researchers. 
    \item We design a sexual violence report annotation and characterization scheme by following the guidelines of Center for Disease Control and Prevention (CDC) and use Amazon Mechanical Turk to annotate about $18,000$ tweets.
    \item We define the sexual violence report tracking as multiple multi-class supervised classification problem and sequence labeling problem (Section~\ref{sec:problem_formulation}, ~\ref{sec:model-development}). To the best of our knowledge, this is the first data-driven, AI based solution towards automated sexual violence report tracking. 
    \item We present two cases by deploying the tracking models over the \#MeToo corpus and a Post-MeToo corpus (Sections \ref{sec:metoo}, \ref{sec:post_metoo}). Our analysis reveals several relationships between violence and perpetrator category.
\end{enumerate}
\section{Literature Review}
\textbf{Barriers to Reporting}: Knowledge about sexual violence is limited by a widespread silence around it. There are different factors working behind this silence. One dominant factor, especially operational in the western world, is the fear of retaliation which is defined by the victim's fear of being ignored or the incident being trivialized by the listeners~\cite{bergman2002reasonableness}. This silence is also often created by the lack of hope in the victim who do not expect the listeners to help them~\cite{banyard2011will,garrett2019understanding}. Furthermore, in many workplaces, sexual violence is often considered as an expression of masculinity, and being a victim is seen as a sign of weakness. This holds women back from reporting any sexual violence happened to them~\cite{fitzgerald1993sexual}. Speaking out is even harder for women in many places in the south Asia. In most places in the Indian subcontinent, for example, ``sex" is a tabooed topic, and talking about sexual harassment is seen as an act of immodesty~\cite{nair2000question}. This challenge is compounded by the strong patriarchal culture~\cite{houston1991speaking} in India and the middle-east~\cite{shalhoub2003reexamining}, where women often depend on men for most things needed for their living including foods, shelter, education, transportation, and health~\cite{mitter1992dependency}. For these and many other reasons, women often remain silent about the violence they experience; and their silence creates a gap in our knowledge about this particular problem. 

\textbf{Current Reporting Mechanisms}: Feminist activists, researchers, artists, and scholars have long been trying to break this silence of women about sexual harassment. There have been many demonstrations on the street and on the news, radio, and television media by the feminist activists that depict the severity of sexual harassment, and the gender-politics related to it~\cite{hester1996women}. With the advancement of information and communication technologies, people have now many effective means of communication including mobile phone and internet. Researchers have tried to leverage them to design different communication platforms for women to report abuses, get safety information, combat harassment, and share their opinions with others. For example, \textit{Hollaback!} is a smartphone application that allows the users to take pictures of the harassers and post the pictures on social media to shame them~\cite{dimond2013hollaback}. Victims can also report harassment through \textit{Harassmap} (harassmap.org), \textit{Bijoya} (bijoya.crowdmap.com), or \textit{Akshara} (akshara.crowdmap.com), applications that show an interactive map of sexual harassment incident places to the users~\cite{young2014harassmap}. One limitation of these systems is, unless the victims are aware of these, the systems won't sustain. For example, \textit{Bijoya} had the latest report in 2013 and \textit{Akshara} had the latest report in 2016. Once the popularity of such a system fades out, it no longer tracks reports of sexual harassment. \textit{Safetipin} is a mobile phone application with which users can mark the unsafe places, and it also informs a user if they enter into an unsafe place. On the other hand, \textit{Circle of 6} (www.circleof6app.com) helps users find other trusted women to accompany them while traveling. \textit{Protibadi}, a mobile application was deployed that allowed women to anonymously report harassment, provided an interactive harassment map, and allowed the users to participate in a discussion around harassment anonymously~\cite{ahmed2014protibadi}. All these applications have been proven to be effective to some extent, and could help some women break the silence and raise their voice against sexual harassment. However, as it has been seen, many women do not use these platforms to talk about sexual harassment because of many larger social, cultural, and political challenges that just cannot be fully mitigated only by easy, accessible, ubiquitous, and intelligent applications and communication tools~\cite{ahmed2016computing}.

\textbf{Promise of Social Media Based Report Tracking}: The \#MeToo movement on Twitter and other social media is considered revolutionary~\cite{manikonda2018twitter}. Such voluntary and spontaneous participation of millions of women across the globe in breaking down a silence that had long been suppressing them was unprecedented in the history~\cite{kearl2018facts}. Through the tweets of these women, we could know information about sexual harassment that was never disclosed before~\cite{rodino2018me}. We could also learn the context, emotion, and repercussions around sexual harassment which would be hard to get otherwise~\cite{khatua2018sounds,bhattacharyya2018metoo}. We argue that the tweets shared by the victims over Twitter have not only helped them release their bottled up emotions, but also have provided us with important data that we can use for better understanding the nature of sexual harassment~\cite{tambe2018reckoning}. Such information can be used for improving existing law, policy, education, and technology to reduce the occurrences of sexual harassment and to extend a better support to the victims~\cite{jaffe2018collective,tippett2018legal}. Moreover, our analysis finds that victims take to social media to report violence experience even in the absence of a social movement. So, we argue that the insights extracted from the \#MeToo movement data can potentially assist in tracking sexual violence reports from social media. 

\section{Data Preparation}
\label{sec:data-preparation}

\subsection{Collection}
\label{sec:data-collection}
We collected about a million tweets containing the \emph{\#MeToo} hashtag from all over the world within the October $15-31$, $2017$ period using the Twitter API. This is the period when the movement was gaining momentum. We have made a $25\%$ random sample of the whole dataset publicly available for other researchers~\footnote{Link is not provided due to the anonymous submission policy.}. We removed the duplicate tweets from the dataset by matching the permanent URLs of the tweets. After that, we removed the re-tweets of original tweets. These cleaning steps reduced the collection size to $520,761$ tweets that were tweeted by $363,090$ unique users. As the tweets are gathered from all over the world, we find that they are written in multiple languages. We translated the non-English tweets to English using the Google Translate API \footnote{https://cloud.google.com/translate/docs/}. 


\subsection{Processing}
\label{sec:data-process}
Our primary inspection finds that the dataset contains tweets that are pertinent to the \#MeToo movement and also some tweets that were irrelevant such as advertisements, spams, etc. Moreover, the pertinent tweets were also of varying nature. For instance, some tweets contained sexual violence reports, some expressed support to the spirit of the \#MeToo movement, and some were opposing or questioning the movement.
We used the Amazon Mechanical Turk [AMT] (details in Section \ref{sec:data-annotation}) to separate the sexual violence reports from the rest of the dataset. However, as the number of tweets is large and the AMT service is not cheap, we wanted to filter the dataset and maximize the chance of getting positive samples. We processed the data as follows.

\textbf{Identify Common Violence Related Verbs}
\label{sec:verb}
We used NLTK~\footnote{https://www.nltk.org/} \textit{part of speech (POS)} tagger to identify the verbs that are present in the tweets. Then, we manually inspected the top-$1000$ most frequent verbs and identified a set of $27$ verbs, $\mathcal{H}$, which are generally associated with sexual violence or events of that nature. The verbs are-- \emph{Abuse, Assault, Attack, Beat, Bully, Catcall, Flirt, Fondle, Force, Fuck, Grab, Grope, Harass, Hit, Hurt, Kiss, Masturbate, Molest, Pull, Rape, Rub, Slap, Stalk, Threat, Touch, Use, Whistle}. These verbs and their variants (e.g., Abuse: abuse, abused, abusing) were present in $103,083$ tweets ($20\%$ of $520,761$).

\textbf{Identify Potential Sexual Violence Reports}
We understand that the presence of a sexual violence related verb in a tweet does not necessarily mean that the tweet reports a sexual violence. To identify potential sexual violence reports, we applied \textit{Semantic Role Labeling (SRL)}~\cite{carreras2005introduction}. Given a sentence, SRL can identify the verb, the agent of the verb, and the details. For example, SRL extracts the tuple \emph{(a jerk, grab, my vagina)} from the tweet \emph{``A jerk grab my vagina at a night club in NYC''}. We used a deep highway BiLSTM with constrained decoding based SRL detection technique ~\cite{he2017deep} that achieves $83.2\%$  F-1  on  the
CoNLL  2005 ~\cite{carreras2005introduction}  test  set  and  $83.4\%$  F-1  on
CoNLL 2012 ~\cite{pradhan-etal-conll-st-2012-ontonotes}. We considered the tweets where SRL finds a tuple $<agent, verb, detail>$ such that,\\
\tab $verb \in \mathcal{H}$ \\ 
\tab $agent$ $=$ \textit{a string with a noun or $3^{rd}$ person pronoun}\\ 
\tab $detail$ $\neq$  \textit{an empty string}

It should be noted that the SRL detection technique identifies presence of negation of an action (e.g., \emph{He did\textbf{n't} kiss her}) or modal of an action (e.g., \emph{Women \textbf{can} abuse men}). We omitted the tweets containing negation or modal from our consideration. In total, we identified $17,935$ ($17.39\%$ of $103,083$) such tweets through this process. We argue that these tweets are more likely to contain sexual violence reports than the tweets where SRL doesn't find such a tuple.

We take a random sample of $16,000$ from these $17,935$ tweets and a random sample of $10,000$ from the rest of the tweets that-- \textbf{i)} didn't pass the SRL filter, thereby don't necessarily contain a violence related verb, and \textbf{ii)} contained at least $5$ words.  In this way, we created a set of $26,000$ tweets to annotate using AMT.

\subsection{Annotation Scheme}
\label{sec:data-annotation}
We prepare the annotation scheme by following the sexual violence reporting guidelines and categorizations provided by the Centers for Disease Control and Prevention (CDC)~\cite{cdc}, the leading national public health institute of the United States. Given a tweet, an AMT worker identifies the following five information-

\begin{enumerate}
    \item \label{Q1} Category of sexual violence, if reported in the tweet. The categories are-
    \begin{enumerate}
        \item (PEN) Completed or attempted forced penetration.
        \item (PEN) Completed or attempted alcohol/drug facilitated penetration.
        \item (PEN) Completed or attempted forced acts in which a victim is made to penetrate someone.
        \item (PEN) Completed or attempted alcohol/drug facilitated acts in which a victim is made to penetrate someone.
        \item (PEN) Nonphysically forced penetration which occurs after a person is pressured to consent. 
        \item (USC) Unwanted sexual contact. 
        \item (NSE) Non-contact unwanted sexual experiences. 
        \item (OTH) A sexual violence incident that doesn't belong to any of the above category.
        \item (NON) The tweet doesn't report a sexual violence.
    \end{enumerate}
    \item \label{Q2} Identity of the victim in case the tweet reports a violence. Possible categories are- (a) (SLF) The tweeter is the victim- in other words, the victim is self-reporting, (b) (nSLF) The tweeter is not the victim.
    \item \label{Q3} Gender of the victim. The options are- (FEM) Female, (MAL) Male, and (UNS) Unspecified.
    \item \label{Q4} Category of the perpetrator (person who inflicts the violence), if specified in the tweet. The categories are-- (INT) Intimate Partner, (FAM) Family Member, (POW) Person in Position of Power, Authority or Trust, (FRN) Friend or Acquaintance, (STR) Stranger, and (PNM) Perpetrator not mentioned or not specific enough.
    \item \label{Q5} Words from the tweet that describe the perpetrator.
\end{enumerate}

\noindent
\textbf{Example 1.} For example, given the following tweet- ``\textit{an old guy at wendy's moved me aside in line by grabbing me by the hips}'', possible annotations would be \emph{\ref{Q1}. USC}, \emph{\ref{Q2}. SLF}, \emph{\ref{Q3}. FEM}, and \emph{\ref{Q4}. STR}, and \emph{\ref{Q5}. ``an old guy''}.

\subsection{Annotated Data}
\label{sec:annotated_data}
Each tweet was annotated by one annotator. Due to the large scale of the data, we couldn't afford multiple annotators for a tweet. However, a manual analysis of the annotated data found the annotations to be of good quality. In total, $18,624$ tweets were annotated by AMT out of $26,000$\footnote{The annotation task is paused at this moment to preserve anonymity of the authors. It will be resumed after this manuscript's reviews are available.}. Table~\ref{tab:sexualviolencedist} shows distributions of the sexual violence categories. In total, there are $11,671$ tweets containing a sexual violence reports and $6,953$ tweets do not contain any sexual violence reports. There are $3,529$ penetration (PEN) related reports that comprise of the first five sexual violence categories. In our sexual violence characterization task (details in Section~\ref{sec:characterization}), we treat the penetration related violence as a whole rather than treating each of the PEN categories separately due to limited sample size for these categories except for the first one. The most frequent category is the \textit{Unwanted Sexual Contact (USC)}.

About $79.13\%$ ($9,236$) of the reports are from the $16K$ SRL filtered data. This demonstrates the effectiveness of the SRL based filtering to maximize getting sexual violence reports. About $20.86\%$ ($2,435$) of the reports came from $10K$ data. So, our dataset also contains a good number of reports which don't follow the SRL based patterns. 

\begin{table*}[t!]
\centering
\resizebox{\linewidth}{!}{%
\begin{tabular}{l|l|l|r}
\toprule
Category & Count & Code & Count \\
\midrule
Completed or attempted forced penetration                                                           & 1919 & \multirow{5}{*}{PEN} & \multirow{5}{*}{3529} \\
Completed or attempted alcohol or drug facilitated penetration                                         & 438  &                      &                       \\
Completed or attempted forced acts in which a victim is made to penetrate someone                   & 699  &                      &                       \\
Completed or attempted alcohol or drug facilitated acts in which a victim is made to penetrate someone & 241  &                      &                       \\
Nonphysically forced penetration which occurs after a person is pressured to consent                & 232  &                      &                       \\
\midrule
Unwanted sexual contact                                                                             & 3980 & USC                  & 3980                  \\
\midrule
Non-contact unwanted sexual experiences                                                              & 1492 & NSE                  & 1492                  \\
\midrule
A sexual violence incident that doesn't belong to any of the above category                         & 2670 & OTH                  & 2670                  \\
\midrule
This tweet doesn't report a sexual violence                                                         & 6953 & NON                  & 6953                  \\
\bottomrule
\end{tabular}%
}
\caption{Sexual violence category distribution in the annotated data}
\label{tab:sexualviolencedist}
\end{table*}

Among the $11,671$ sexual violence reports, $7,395$ ($63.36\%$) were self-reported (SLF) by the victims and the remaining $4,276$ were not self reported (nSLF). Our analysis finds that the nSLF set contains reports where the tweeter is reporting a sexual violence of other victim such as a family member or an acquaintance. It also has reports where the victim is mentioned but not specifically enough to understand the identity.

Question~\ref{Q3} asked about gender identity of the victim. The percentage of FEM, MAL, and UNS were $41.43\%$, $9.39\%$, and $49.18\%$, respectively. A gender-violence bi-variate analysis reveals that females face more unwanted sexual contact (USC) violence than what males face. Table~\ref{tab:gender_violence} shows frequency (percentage) of each gender category for each violence category. Percentage of USC among FEM is $38.02\%$ whereas it is $26.56\%$ among MAL. Also, USC is the most common violence in FEM whereas for MAL it is PEN. 

\begin{table}[h!]
\centering
\resizebox{\linewidth}{!}{%
\begin{tabular}{l|llll}
\toprule 
 & NSE   & OTH   & PEN   & USC   \\
\midrule
FEM & 610 (14.12) & 843 (19.51) & 1225 (28.35) & 1643 (38.02) \\
MAL & 90 (9.19)  & 197 (20.12) & 432 (44.13) & 260 (26.56) \\
UNS & 679 (13.24) & 1264 (24.65) & 1246 (24.30) & 1939 (37.81) \\
\bottomrule
\end{tabular}%
}
\caption{Frequency (Percentage) of each Gender-Violence combination}
\label{tab:gender_violence}
\end{table}

Question~\ref{Q4} asked about perpetrator's nature. We find that $7,509$ ($64.34\%$) of the sexual violence reports contain clear categorization of the relationship between the victim and the perpetrator. The relationship can be one of the first 5 categories described in Question~\ref{Q4}. Table~\ref{tab:perpetratorcategory} shows distribution of the categories. The majority of the perpetrators are stranger to the victims. The second major category consists of perpetrator from person in position of power or authority or trust. There are $4,162$ ($35.66\%$) violence reports which either don't mention the relationship specifically or don't mention about perpetrator at all (PNM category of Question~\ref{Q4}).

\begin{table}[h!]
    \centering
    \resizebox{\linewidth}{!}{%
    \begin{tabular}{l|r||l|r||l|r}
        \toprule
        Text    & Count & Text  & Count  & Text  & Count\\
        \midrule
        he      & 509 & a boy      & 69 & my boss           &47\\
        men     & 276 & boss       & 64 & they              &42\\
        a man   & 269 & women      & 63 & stranger          &42\\
        a guy   & 192 & friend     & 60 & a woman           &40\\
        man     & 114 & a stranger & 56 & boyfriend         &35\\
        someone & 77  & the man    & 55 & she               &31\\
        guy     & 75  & the guy    & 49 & manager           &29\\
        \bottomrule
    \end{tabular}%
    }
    \caption{Common perpetrator words}
    \label{tab:ppt_common_words}
\end{table}

Question~\ref{Q5} requires the AMT coders to highlight the words of a tweet that describe the perpetrator of a sexual violence. We took several precautionary measures to ensure good quality annotation and discourage spam activities. For example, we discarded the data where coders responded (i) NON to Question~\ref{Q1} or (f) PNM to Question~\ref{Q4} but at the same time highlighted words in Question~\ref{Q5}. We also alerted the coders when they deliberately or mistakenly highlighted the whole tweet. Our analysis finds that the highlighted words by and large comprise of nouns such as \emph{boss}, \emph{teacher}, \emph{uncle}, etc. or pronouns such as \emph{he}, \emph{she}. Table~\ref{tab:ppt_common_words} shows some of the most frequent perpetrator indicating texts that were highlighted by the coders. The table has both `a man' and `man'. It could be because a tweeter wrote without the article `a' or because a coder didn't highlight the article. The average number of highlighted words in the tweets is $2.19$. We use these words to train a sequence labeling model that can predict perpetrator of sexual violence (details in Section~\ref{sec:ppc}).

\begin{table}[h!]
\centering
\resizebox{\linewidth}{!}{%
\begin{tabular}{l|l|r}
\toprule
Category & Count & \% \\
\midrule
(INT) Intimate Partner & 545  & 7.26  \\
(FAM) Family Member (not Intimate Partner) & 837  & 11.14  \\
(POW) Person in Position of Power/Authority/Trust & 2045 & 27.23 \\
(FRN) Friend or Acquaintance & 1735 & 23.11 \\
(STR) Stranger & 2347 & 31.26 \\
\midrule
Total & 7509 & 100 \\
\bottomrule
\end{tabular}%
}
\caption{Perpetrator category distribution}
\label{tab:perpetratorcategory}
\end{table}

\section{Problem Formulation}
\label{sec:problem_formulation}
We design the sexual violence report tracking as a two step process- \textbf{i)} detecting sexual violence reports and \textbf{ii)} characterizing the reports with respect to violence nature, victim's identity, gender identity, and relationship with perpetrator.


\subsection{Sexual Violence Report Detection}
\label{sec:detection}
We define the sexual violence report detection task as a supervised binary classification problem where the set of classes, $\mathcal{C} =  \{SVR, nSVR\}$. Formally, given $\mathcal{X}$, a set of all tweets, and a training set $\mathcal{T}$ of annotated tweets  $\langle t, c\rangle$, where  $\langle t, c\rangle \in \mathcal{X} \times \mathcal{C}$, our goal is to learn a function $\gamma$ such that $\gamma : \mathcal{X} \rightarrow \mathcal{C}$, in other words, it maps tweets to $\{SVR, nSVR\}$.

\subsection{Sexual Violence Report Characterization}
\label{sec:characterization}
We formulate the characterization of reports as multiple multi-class supervised classification problems. Formally, given $\mathcal{X_{SVR}}$, a set of all tweets that have been mapped to $\{SVR\}$ by $\gamma$, and a training set $\mathcal{T_{SVR}}$ of annotated tweets $\langle t, (c, v, g, p)\rangle$ where $t \in \mathcal{X_{SVR}}$, $c \in \{PEN, USC, NSE, OTH\}$, $v \in \{SLF, nSLF\}$, $g \in \{FEM, MAL, UNS\}$, $p \in \{INT, FAM, POW, FRN, STR, PNM\}$, our goal is to learn functions $\zeta, \upsilon, \xi, \rho$ such that-\\
\tab $\zeta : \mathcal{X_{SVR}} \rightarrow \{PEN, USC, NSE, OTH\}$\\
\tab $\upsilon : \mathcal{X_{SVR}} \rightarrow \{SLF, nSLF\}$\\
\tab $\xi : \mathcal{X_{SVR}} \rightarrow \{FEM, MAL, UNS\}$\\
\tab $\rho$ : $\mathcal{X_{SVR}}$ $\rightarrow$ $\{INT, FAM, POW, FRN, STR, PNM\}$\\

In addition, we model the perpetrator tagging task as a sequence labeling problem. Specifically, given a tweet $t$, the goal is to identify the consecutive set of words in $t$ that denotes the perpetrator who caused the sexual violence. 

\section{Model Development}
\label{sec:model-development}
In the following sections, we describe the processing of the annotated data, modeling of the problems, and comparison of performances of multiple learning techniques.

\subsection{Data Processing}
We preprocessed the tweets following a series of steps. First, we converted the tweets to all lowercase. Then, a Python package named tweet-preprocessor~\footnote{https://pypi.org/project/tweet-preprocessor/} was used to remove URLs, emojis, and smileys from the tweets. We also removed the hashtag \#metoo from the tweets. After that, these cleaned tweets were transformed to vectors using Bag-of-Words (BOW) model and TF-IDF (with sublinear tf). We used unigrams and bigrams as features (ngrams with n = 1, 2). There were $147,370$ such features. The ngrams that were present in only one tweet (very rare) or present in more than $25\%$ of all tweets (very common) were discarded from feature set. This step reduced the number of features to $34,688$. From these, we selected the most frequent $5,000$ ngram features and experimented with these in all the experiments.

\textbf{Note:} We experimented with stopwords removal and stemming. However, these steps decreased the overall performance. So, we decided not to use these two steps.


\subsection{Model Selection}
We experimented with multiple supervised learning algorithms. Specifically, we used Support Vector Machine (SVM) with linear kernel, Random Forest (RDF) with 100 estimators, and Gradient Boosting (GDB). In some cases, the SVM and GDB outperformed others in terms of f1-measure. However, we observed a better performance when all of the three classifiers were combined to build a majority voting classifier (MVC). We used a Python package named scikit-learn to build MVC. It predicts the class label based on the \textit{argmax} of the sums of the predicted probabilities by SVM, RDF, and GDB. Table~\ref{tab:performance} shows performances of the classifiers in terms of precision (P), recall (R), and F1-measure(F1). We conducted $5$-fold cross-validation. In each fold, $80\%$ of the data was used for training and the remaining $20\%$ was used for testing. Then took the average of the performances of the folds. The highest F1-measure for each category is highlighted in the table. MVC achieved the highest F1-measures in $10$ of the $17$ total categories.

\textbf{Note:} We also experimented with word embedding based techniques (GLOVE~\cite{pennington2014glove}) with deep learning models (LSTM, IndRNN~\cite{li2018independently}). However, we didn't observe better performance, particularly in task~\ref{sec:characterization}. This could be because of the fact that the sample size was small to learn the model parameters.

\subsection{Sexual Violence Report Detection}
Recall that the dataset has about 11K sexual violence reports and 7K other tweets. MVC classifier achieves the best performance in detecting sexual violence reports in tweets. It can identify a report with a precision, recall, and F1-measure of $0.804$, $0.834$, and $0.808$, respectively. To further understand the effectiveness of MVC, we used its $SVR$ class probabilities to conduct ranking evaluation. We measured the accuracy of the top-$k$ sentences by two commonly used measures- Precision-at-k (P@k), and Average Precision-at-k (AvgP@k). Table~\ref{tab:svr_ranking} shows these measures for various value of $k$. In general, MVC achieves excellent performance in ranking. For top 1000 sentences, its precision is $0.951$. This indicates that MVC has a strong agreement with human annotators on detecting sexual violence reports.

\begin{table}[!htb]
  \centering
  \resizebox{\linewidth}{!}{%
    \begin{tabular}{@{}l|rrrrrrr@{}}
    \toprule
    k      & 25 & 50 & 100 & 300   & 500   & 1000  & 2500  \\
    \midrule
    P@k    & 1.0  & 1.0  & 1.0   & 0.993 & 0.982 & 0.951 & 0.798 \\
    \midrule
    AvgP@k & 1.0  & 1.0  & 1.0   & 0.997 & 0.993 & 0.950 & 0.935 \\
    \bottomrule
    \end{tabular}%
    }
    \caption{Ranking evaluation of sexual violence reports}
    \label{tab:svr_ranking}
\end{table}

        

\subsection{Sexual Violence Report Characterization}
In general, MVC outperforms other classifiers in characterizing sexual violence reports. It achieves the highest weighted-average F1-measure in all the categorization tasks. The weighted-average F1-measure for violence, victim, gender, and perpetrator categorization is $0.57$, $0.82$, $0.612$, and $0.55$, respectively.

\subsubsection{Violence and Victim Category}
MVC also achieves the best precision and recall in detecting penetration related ($PEN$) sexual violence reports. In case of detecting unwanted sexual contact ($USC$), the GDB has the best precision ($0.736$) and F1-measure ($0.724$). None of the classifiers performed well in case of detecting non-contact sexual experience ($NSE$). It could be because of the comparatively limited sample or class imbalance. The SVM achieves the best F1-measure ($0.4$) in this category which is $0.15$ higher than a random guess. SVM and MVC comparatively perform well in detecting whether the sexual violence report is a self-report or not. MVC achieves the best weighted average F1-measure- $0.82$.

\subsubsection{Gender Category}
Similar argument can be made for $MAL$ category in gender categorization. Although the performances are reasonable for $FEM$, and $UNS$ but $MAL$ measures, particularly recall and F1-measure, are lower than random guess. This suggests that many $MAL$ reports were mis-classified into other categories. A further look into the confusion matrix reveals that $57\%$ and $35\%$ of $MAL$ reports were classified as $UNS$ and $FEM$, respectively.

\subsubsection{Perpetrator Category}
\label{sec:ppc}
MVC achieves the best weighted F-1 measure in all perpetrator categories. The weighted average F1-measure is $0.55$, $0.38$ higher than a random guess.

Our annotation scheme collects the perpetrator denoting words from the tweets (see Section~\ref{sec:data-annotation}, \ref{sec:annotated_data} and Example 1 for reference). We used Conditional Random Field (CRF)~\cite{lafferty2001conditional} to train a sequence labeling model using these words. We used the \textbf{BIE} (begin-inside-end) notation to tag the perpetrator word sequences. Words those were not part of the perpetrator sequence, were assigned the \textbf{O} tag. For instance, the tweet ``in the subway an old man grabbed me'' would be tagged as- \{in-O, the-O, subway-O, an-B, old-I, man-E, grabbed-O, me-O\}. We used the words, their prefix and suffix, and their parts-of-speech tags as features while training the sequence labeling model. In total, $8,614$ sexual violence reporting tweets were labeled with perpetrator words. We did a $5$-fold cross-validation. Table~\ref{tab:ppt_perf} shows performance of the CRF perpetrator text labeling model in terms of precision, recall, and F1-measure. The model achieves a weighted-average F1-measure of $0.943$ ($0.70$ if class \textbf{O} is not considered). We observe that the performance for class \textbf{I} is low. As most of the highlighted perpetrator text ($6,653$ out of $8,614$) was less than or equal two words, we didn't have enough sample for class \textbf{I}. For the \textbf{B} and \textbf{E} classes, the F1-measures were above $0.76$.

\begin{table}[h!]
\centering
\begin{tabular}{@{}l|lllr@{}}
\toprule
Class            & Precision & Recall & F1-measure & Support \\ \midrule
O           & 0.959     & 0.989  & 0.974    & 35440   \\
B       & 0.840     & 0.703  & 0.765    & 1723    \\
E       & 0.883     & 0.739  & 0.805    & 1723    \\
I       & 0.601     & 0.218  & 0.320    & 871     \\
Avg / total & 0.943     & 0.949  & 0.943    & 39757   \\ \bottomrule
\end{tabular}
\caption{Performance of perpetrator labeling}
\label{tab:ppt_perf}
\end{table}



\begin{table*}[t!]
\centering
\resizebox{\linewidth}{!}{%
\begin{tabular}{@{}l|l|lll|lll|lll|lll@{}}
\toprule
\multirow{2}{*}{Classifier}             & \multirow{2}{*}{Categories} & \multicolumn{3}{c}{SVM}        & \multicolumn{3}{c}{GDB}        & \multicolumn{3}{c}{RDF} & \multicolumn{3}{c}{MVC}        \\ \cmidrule(l){3-14} 
                                        &                        & P     & R     & F-1            & P     & R     & F-1            & P      & R      & F-1   & P     & R     & F-1            \\
\midrule
\multirow{3}{*}{Sexual Violence Report} & $nSVR$                   & 0.708 & 0.666 & 0.662          & 0.710 & 0.620 & 0.632          & 0.736  & 0.628  & 0.648 & 0.752 & 0.638 & \textbf{0.656} \\
                                        & $SVR$                    & 0.810 & 0.802 & 0.798          & 0.786 & 0.810 & 0.792          & 0.798  & 0.830  & 0.804 & 0.804 & 0.834 & \textbf{0.808} \\
                                        & weighted-Avg                    & 0.772 & 0.754 & 0.744          & 0.758 & 0.740 & 0.732          & 0.778  & 0.756  & 0.746 & 0.782 & 0.760 & \textbf{0.750} \\
\midrule
\multirow{5}{*}{Violence Category}      & $NSE$                    & 0.456 & 0.358 & \textbf{0.400} & 0.444 & 0.358 & 0.394          & 0.446  & 0.276  & 0.340 & 0.484 & 0.336 & 0.394          \\
                                        & $OTH$                    & 0.422 & 0.452 & 0.432          & 0.434 & 0.454 & \textbf{0.440} & 0.434  & 0.420  & 0.426 & 0.444 & 0.434 & 0.438          \\
                                        & $PEN$                    & 0.560 & 0.532 & 0.546          & 0.548 & 0.576 & 0.560          & 0.574  & 0.576  & 0.576 & 0.578 & 0.584 & \textbf{0.580} \\
                                        & $USC$                    & 0.678 & 0.718 & 0.694          & 0.736 & 0.720 & \textbf{0.724} & 0.664  & 0.760  & 0.708 & 0.686 & 0.758 & 0.718          \\
                                        & weighted-Avg                    & 0.558 & 0.554 & 0.554          & 0.572 & 0.570 & 0.566          & 0.556  & 0.566  & 0.556 & 0.572 & 0.580 & \textbf{0.570} \\
\midrule
\multirow{3}{*}{Victim Category}        & $SLF$                    & 0.780 & 0.714 & \textbf{0.744} & 0.766 & 0.532 & 0.626          & 0.77   & 0.694  & 0.726 & 0.802 & 0.688 & 0.738          \\
                                        & $nSLF$                   & 0.842 & 0.878 & 0.860          & 0.768 & 0.902 & 0.832          & 0.83   & 0.876  & 0.852 & 0.834 & 0.898 & \textbf{0.864} \\
                                        & weighted-Avg                    & 0.818 & 0.818 & 0.818          & 0.766 & 0.770 & 0.756          & 0.81   & 0.808  & 0.802 & 0.826 & 0.822 & \textbf{0.820} \\
\midrule
\multirow{4}{*}{Gender Category}        & $FEM$                    & 0.618 & 0.538 & 0.576          & 0.708 & 0.476 & 0.570          & 0.648  & 0.540  & 0.586 & 0.680 & 0.522 & \textbf{0.590} \\
                                        & $MAL$                    & 0.642 & 0.092 & \textbf{0.154} & 0.616 & 0.078 & 0.138          & 0.570  & 0.062  & 0.108 & 0.642 & 0.074 & 0.130          \\
                                        & $UNS$                    & 0.618 & 0.782 & 0.688          & 0.614 & 0.892 & \textbf{0.728} & 0.632  & 0.826  & 0.716 & 0.624 & 0.858 & 0.722          \\
                                        & weighted-Avg                    & 0.618 & 0.616 & 0.592          & 0.654 & 0.642 & 0.606          & 0.630  & 0.634  & 0.604 & 0.650 & 0.642 & \textbf{0.612} \\
\midrule
\multirow{7}{*}{Perpetrator Category}   & $FAM$                    & 0.692 & 0.466 & 0.552          & 0.696 & 0.472 & 0.552          & 0.674  & 0.436  & 0.526 & 0.718 & 0.482 & \textbf{0.568} \\
                                        & $FRN$                    & 0.512 & 0.346 & 0.410          & 0.582 & 0.326 & 0.418          & 0.618  & 0.308  & 0.406 & 0.610 & 0.330 & \textbf{0.426} \\
                                        & $INT$                    & 0.708 & 0.340 & 0.452          & 0.546 & 0.354 & 0.424          & 0.640  & 0.314  & 0.414 & 0.682 & 0.360 & \textbf{0.462} \\
                                        & $PNM$                    & 0.502 & 0.770 & 0.604          & 0.488 & 0.814 & 0.610          & 0.486  & 0.808  & 0.608 & 0.500 & 0.830 & \textbf{0.622} \\
                                        & $POW$                    & 0.656 & 0.482 & 0.552          & 0.726 & 0.426 & 0.536          & 0.660  & 0.448  & 0.526 & 0.702 & 0.478 & \textbf{0.564} \\
                                        & $STR$                    & 0.574 & 0.434 & 0.486          & 0.622 & 0.426 & 0.496          & 0.590  & 0.428  & 0.490 & 0.642 & 0.432 & \textbf{0.508} \\
                                        & weighted-Avg                    & 0.568 & 0.548 & 0.532          & 0.590 & 0.550 & 0.534          & 0.578  & 0.546  & 0.524 & 0.604 & 0.564 & \textbf{0.550} \\
\bottomrule
\end{tabular}%
}
\caption{Performance of different classifiers on the detection and characterization tasks}
\label{tab:performance}
\end{table*}
\vspace{-0.3in}
\section{Case Studies}
Using the above described classifiers, we conduct an exploratory analysis on two cases- \textbf{i)} the \#MeToo movement and \textbf{ii)} the Post-Metoo era. The data collection method for the \#Metoo has been explained in Section~\ref{sec:data-collection}. For the Post-MeToo case, we collected $631,677$ tweets from May, 2019 when the \#MeToo movement was not the major news topic in the world.

\subsection{The \#MeToo Movement}
\label{sec:metoo}
We apply our sexual violence report detecting and characterizing classifiers on the $0.49$ million tweets (except the tweets which were used to train the models) which were collected during the \#MeToo movement's prime time (see details in Section~\ref{sec:data-collection}). Our goal is to understand to what extent sexual violence experiences were reported during this movement, what were the nature of the violence, who are the perpetrators, and more.

\subsubsection{Extent of Sexual Violence} Our classifiers identify $106,885$ tweets containing sexual violence reports. $59,776$ of them are self-reported ($SLF$). To conduct a robust analysis with high-probable violence reports, we identified the $SLF$s with $SVR$ class probability greater than or equal to $0.7$. This threshold is set empirically, after attempting with multiple smaller and larger values. There are $25,163$ self-reports which surpass this threshold. We continue our analysis with these high $SVR$ class probability, $SLF$ labeled reports only. $4,462$ of these $SLF$ reports are from female ($FEM$) victims, $105$ from male ($MAL$), and $20,596$ belong to $UNS$. 

\subsubsection{Nature of Violence}
Our violence category classifier finds that out of $25,163$ $SVR$, $SLF$ labeled reports, $1,866$, $3,477$, $9,193$, and $10,627$ reports are classified as $NSE$, $OTH$, $PEN$, and $USC$, respectively.

\begin{table}[h!]
\centering
\resizebox{\linewidth}{!}{%
\begin{tabular}{@{}l|ll|lr@{}}
\toprule
Category & Frequency & Percentage & Tagged & Percentage \\ \midrule
FAM      & 1112      & 10.39      & 442    & 39.75      \\
FRN      & 2568      & 24.00      & 731    & 28.47      \\
INT      & 751       & 7.02       & 230    & 30.63      \\
POW      & 2408      & 22.51      & 1865   & 77.45      \\
STR      & 3859      & 36.07      & 939    & 24.33      \\ \bottomrule
\end{tabular}%
}
\caption{Perpetrator category distribution}
\label{tab:ppcm_dist}
\end{table}

\subsubsection{Perpetrator Category}
The perpetrator classifier finds that $10,698$ ($42.5\%$) of the reports contain specific perpetrator information and the rest didn't mention perpetrator in the tweet. Table~\ref{tab:ppcm_dist} shows the distribution of perpetrator categories among these reports. Overall, we observe a large number of violence committed by $STR$ ($36.07\%$). Only $7.02\%$ of the violence was attributed to $INT$.

\subsubsection{Perpetrator Annotation}
\begin{figure*}[t!]
\vspace{-0.4in}
    \centering
    \hspace{-0.75in}
    \begin{subfigure}[t]{0.18\textwidth}
        \centering
        \includegraphics[width=1.3\linewidth]{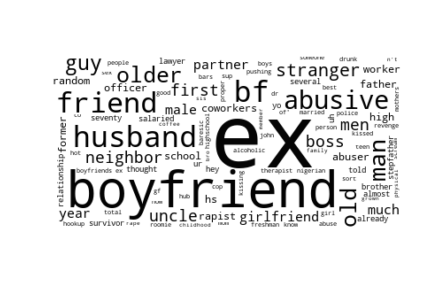}
        \vspace{-0.4in}
        \caption{INT}
    \end{subfigure}
    ~ 
    \begin{subfigure}[t]{0.18\textwidth}
        \centering
        \includegraphics[width=1.3\linewidth]{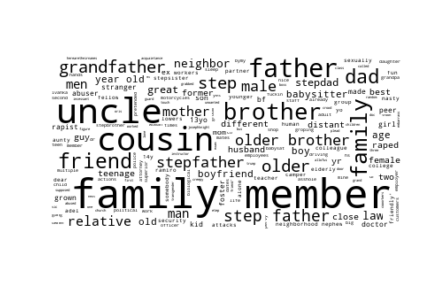}
        \vspace{-0.4in}
        \caption{FAM}
    \end{subfigure}
    ~ 
    \begin{subfigure}[t]{0.18\textwidth}
        \centering
        \includegraphics[width=1.3\linewidth]{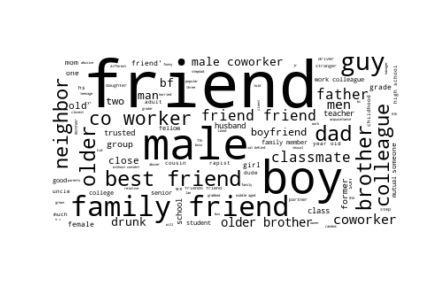}
        \vspace{-0.4in}
        \caption{FRN}
    \end{subfigure}
    ~ 
    \begin{subfigure}[t]{0.18\textwidth}
        \centering
        \includegraphics[width=1.3\linewidth]{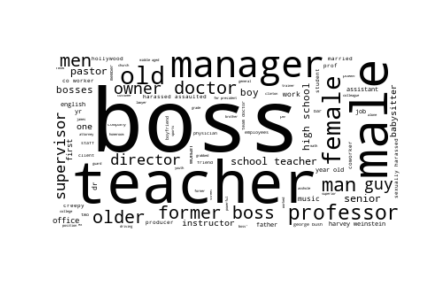}
        \vspace{-0.4in}
        \caption{POW}
    \end{subfigure}
    ~
    \begin{subfigure}[t]{0.18\textwidth}
        \centering
        \includegraphics[width=1.3\linewidth]{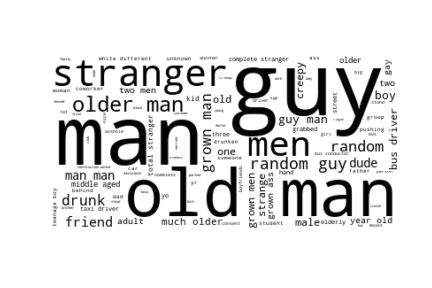}
        \vspace{-0.4in}
        \caption{STR}
    \end{subfigure}%
    \vspace{-0.2in}
    \caption{Perpetrator text tagged by CRF in each perpetrator category. Font size is proportion to relative frequency.}
    \label{fig:perp_wc}
\end{figure*}

We apply the CRF perpetrator text labeling model on the $10,698$ reports where perpetrator information is provided. Overall, the model could tag the perpetrator denoting text in $3,391$ reports. The rightmost two columns in Table~\ref{tab:ppcm_dist} shows the number and percentage of tagged reports in each category. Figure~\ref{fig:perp_wc} shows a wordcloud of CRF tagged perpetrator text for each perpetrator category. The font size signifies relative frequency of a text. Overall, CRF identifies different roles under each category. For example, the $FAM$ cloud shows \emph{father}, \emph{brother}, \emph{cousin}, \emph{uncle}, \emph{step father} etc. The $POW$ cloud shows \emph{boss}, \emph{teacher}, \emph{manager}, \emph{professor} and so on. In Section~\ref{sec:post_metoo}, we provide more examples of reports with annotated perpetrator text. 

\subsubsection{Violence-Perpetrator Relation}
\begin{figure*}[t!]
\centering
\includegraphics[width=\linewidth]{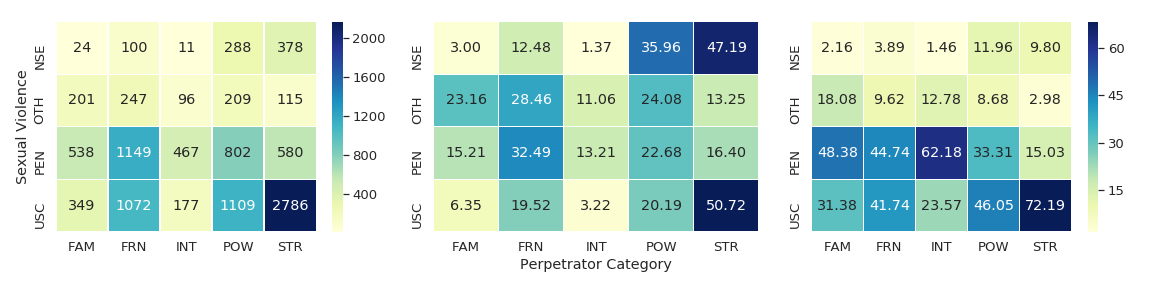}
\caption{Distribution of Sexual Violence and Perpetrator pairs. (\textit{left}: frequency, \textit{middle}: percentage of sexual violence categories across the perpetrators, \textit{right}: percentage of perpetrator categories across the violence.}
\label{fig:svc_ppc}
\end{figure*}

We further investigate the relation between perpetrator category and the nature of violence. Figure~\ref{fig:svc_ppc} shows the number of reports for each violence-perpetrator category pairs (left). Overall, the most common reports are about unwanted sexual contact by strangers ($USC-STR$) and the least common is non-contact unwanted sexual experience by intimate partners ($NSE-INT$). Figure~\ref{fig:svc_ppc} also shows the percentages of sexual violence categories across the perpetrators (middle) and the percentages of perpetrator categories across the violence (right). An interesting observation is that intimate partners mostly conduct penetration related violence ($62.18\%$) but, penetration violence is least done by intimate partners ($13.21\%$). Strangers mostly commit unwanted sexual contacts ($72.19\%$); almost $5$ times more than the second most common violence category ($PEN, 15.03\%$). At the same time, strangers are also the most common perpetrator of the $USC$ category ($50.72\%$); more than twice as common as the next probable perpetrator category ($POW, 20.19\%$). For non-contact unwanted sexual experiences, person in position of power/authority/trust ($POW$) and $STR$ are significantly more responsible than others. Also, $POW$ is the second most common perpetrator class in each of the violence categories. Overall, the perpetrators who are closer to the victims ($FAM$, $INT$, $FRN$) are related to severe sexual violence ($PEN$) more than to what the less closer perpetrators ($STR$, $POW$) relate. Also, the less closer perpetrators ($STR$, $POW$) are more related to mild sexual violence ($USC$, $NSE$) compared to the closer perpetrators ($FAM$, $INT$, $FRN$).


\begin{table*}[ht!]
\centering
\resizebox{\linewidth}{!}{%
\begin{tabular}{l|l|l}
\toprule
No. & $SVC$ & \multicolumn{1}{c}{Tweet}\\
\midrule
1 & PEN & i was 8 years old asleep when my \textbf{{\color{red}biological father}} first raped me. he didn't give a fuck about what i was wearing. instead   \\
2 & USC & \textbf{{\color{red}my boss}} at dominos was telling me about how he was walking his dog at 2am \&; some guy in a truck stopped to talk to \\
3 & USC & my mother was married 12 times and one of the use to touch me i am a \textbf{{\color{red}molested victim}} ...i got saved when i was 14   \\
4 & PEN & i also, zero hates bible thumpers. he hates religion. when \textbf{{\color{red}his pastor}} used to rape him, he would insist it was god  \\
5 & USC & rt @drippinswagu: @parkersuave is it the \textbf{{\color{red}same guy}} that tried to touch dababy ass when he agreed to take the picture                                \\
6 & USC & likethe nigga friend could be dead ass \textbf{{\color{red}struggle dude}} in jail ask him for a couple dollars he say no cus he strugglin    \\
7 & USC & tw// assaulti was wearing leggings and an oversized sweatshirt when a \textbf{{\color{red}16 year old boy}} grabbed my thighs and told       \\
8 & USC & @parkersuave is it the \textbf{{\color{red}same guy}} that tried to touch dababy ass when he agreed to take the picture                                                  \\
9 & USC & @jeff\_lj\_lloyd \textbf{{\color{red}he def}} used his body to push the guy back until he fell, then tried to act like he didn't do anythin \\
10 & PEN & this \textbf{{\color{red}one guy}} told us how his father died and he said dead asss word for word woke up in the middle of the night \\
\bottomrule
\end{tabular}%
}
\caption{Examples of self-reported sexual violence reports. Perpetrator texts are annotated in red.}
\label{tab:examples}
\end{table*}

\vspace{-3mm}
\subsection{Post-MeToo Period}
\label{sec:post_metoo}
\vspace{-3mm}
Even though the spirit of the \#MeToo movement is still present to some extent, the momentum diminished from its initial period. We wanted to investigate- to what extent people take social media to share their sexual violence experience without the presence of an active online movement. So, we collected about $631,677$ tweets from May, 2019, one and a half year after the movement became viral, using the Twitter Streaming API by querying the $27$ keywords in $\mathcal{H}$ (see Section~\ref{sec:verb} for reference). Then, we applied the classification and sequence labeling models over these tweets. In total, we find $4,830$ ($0.76\%$) self-reported sexual violence reports that had a $SVR$ class probability $>= 0.7$. The CRF perpetrator tagger could label $655$ of these reports with perpetrator text. Table~\ref{tab:examples} shows top-$10$ sexual violence reports with respect to the $SVR$ class probability. The CRF tagged perpetrator text in each report is highlighted with red color bold font. The predicted $SVC$ category is also presented on the left column. CRF could correctly tag the complete perpetrator text in $6$ cases (e.g., $1$, $7$). However, there are some false positives as well, $2, 4, 6$ for example. We plan to continue get more data annotated, including tweets without \#MeToo hashtag, and improve the quality of the models which can further reduce the amount of false positives. 



\section{Discussion}
In this paper, we present a data-driven, supervised learning approach of mining sexual violence reports from social media. We envision several pathways regarding how the outcomes of this paper can create impact in the real world. First, the sexual violence supporting coalitions across the world can benefit from this work. Currently, the coalitions depend on the victims' self-reports to them before they can provide any support. An automated tracker can assist the coalitions in identifying the victims. As explained in the Literature Review section, many victims do not go to the authority due to different reasons. An automated tracker can find the self-reported victims in social networks and connect with the authorities. Second, we understand that there is room for adversarial attacks. One way to mitigate those attacks is to design an human-in-the-loop system. The output from this automated tracker can be feed to a set of human domain experts before any decisions or actions are made on the identified reports. Third, due to the sensitivity of the issue, it is possible that the user who publicly shared a sexual violence report on social media may not actually want to be contacted by support centers or law and regulatory entities. It is an issue that demands further research; how to support the victims without the conventional support centers. One possible idea could be creating a social community of sexual violence victims that can provide mental and logistical supports to the peers. We plan to continue working along this line in future.
\section{Limitations}
We present a data-driven, supervised learning approach of mining sexual violence reports from social media in this paper. Our proposed technique can identify the reports with a precision and recall of $80.4\%$ and $83.4\%$, respectively. Our techniques can also characterize the identified reports in terms of violence severity, perpetrator category, and gender. We also automatically tag the perpetrator text if that is present in a report. However, there are some limitations. First, we only used the unigram and bigram features. One possible improvement could be exploiting the user specific and network specific features. These features can particularly improve the gender and victim characterization. Second, the lack of samples in some characterization classes created class imbalance problem. Although, we experimented with undersampling and oversampling methods (details not provided in this paper), but that didn't improve generalization ability of the models. That is why proceeded with actual sample sizes. One can collect more samples of the undersampled classes and improve the overall performances of the models. More data samples would also allow leveraging deep learning based models. Third, we didn't consider credibility of a report into account. So, our models cannot discern a true sexual violence report from a false claim.

\vspace{-3mm}
\section{Conclusion}
In this paper, we address the sexual violence report detection and characterization task using data-driven, machine learning approach. In the process, we make several contributions. First, we collect about a million \#MeToo hashtagged tweets from the period when the movement was in its prime. We systematically annotated a subset of these tweets using Amazon Mechanical Turk following Center for Disease Control and Prevention (CDC) guidelines. To the best of our knowledge, it is the first annotated dataset of sexual violence reports. We employed supervised classification and sequence labeling algorithms to train models using these annotated data. Our models can identify the sexual violence reports with a precision of $80.4\%$ and characterize the reports with respect to four characteristics. Our sequence labeling model can annotate the perpetrator identity from a report with an average F1-measure of $76\%$. We use these models to analyze the full \#MeToo corpus and another Post-MeToo corpus. Our analysis finds some interesting insights such as the relationship between violence severity and perpetrator category. We argue that our contributions in this paper, the annotated data, models, and insights, can be useful in designing an automated, scalable, and reliable sexual violence reports tracking system that can assist the victims by connecting them with appropriate authorities and at the same time help relevant entities such as human rights groups, law and regulatory agencies to reach the victims. In future, our goal is to address the limitations and implement an automated data-driven tracking system.


\textbf{Data Sharing and Reproducibility Statement}
We have made $25\%$ of our collected \#MeToo hashtagged data publicly available. We plan to make the full data public contingent on the decision of the paper. We also plan to make $50\%$ of our annotated data open for researchers. Moreover, we will make our code public through \textit{github} to facilitate reproducibility of our findings.  

\begin{small}
\bibliographystyle{aaai}
\bibliography{main}
\end{small}
\end{document}